\documentclass[preprint,3p,times, twocolumn]{elsarticle}

\usepackage{amssymb}
\usepackage{float} 
\usepackage{dblfloatfix}
\usepackage{amsmath}
\usepackage{graphicx}
\graphicspath{{figure/}}
\usepackage{subfigure}
\biboptions{numbers,sort&compress}
\usepackage{cuted} 
\usepackage{placeins}
\usepackage{afterpage}
\usepackage[colorlinks=true, linkcolor=lightblue, citecolor=lightblue, urlcolor=lightblue]{hyperref}
\usepackage{color}
\usepackage[dvipsnames]{xcolor}
\usepackage[defaultcolor=Maroon]{changes}

\begin{document}

\begin{frontmatter}



\title{Customizing pseudospin unidirectional states of acoustic and electromagnetic waves in two-dimensional phoxonic topological insulators via multi-objective strategies} 


\author[label1]{Gang-Gang Xu}
\author[label2]{Xiao-Shuang Li}
\author[label3]{Tian-Xue Ma\corref{cor1}}
\author[label4,label3]{Xi-Xuan Liu}
\author[label4]{Xiao-Wei Sun}
\author[label1,label3]{Yue-Sheng Wang\corref{cor1}}
\affiliation[label1]{organization={Department of Mechanics, School of Mechanical Engineering, Tianjin University},
            city={Tianjin},
            postcode={300350}, 
            country={China}}
\affiliation[label2]{organization={College of Civil Engineering and Architecture, Hebei University},
	city={Baoding},
	postcode={071002}, 
	country={China}} 	
\affiliation[label3]{organization={Department of Mechanics, School of Physical Science and Engineering, Beijing Jiaotong University},
	city={Beijing},
	postcode={100044}, 
	country={China}} 	
\affiliation[label4]{organization={School of Mathematics and Physics, Lanzhou Jiaotong University},
	city={Lanzhou},
	postcode={730070}, 
	country={China}} 

\cortext[cor1]{Corresponding authors. 
	 \\\textit{Email addresses:} matx@bjtu.edu.cn (T.-X. Ma), yswang@tju.edu.cn (Y.-S. Wang)}

\begin{abstract}
Topological materials for classical waves offer remarkable potential in applications such as sensing, waveguiding and signal processing, leveraging topological protection effects like strong robustness, immunity to backscattering and unidirectional transmission. This work presents the simultaneous inverse design of pseudospin-dependent  topological edge states for acoustic and electromagnetic waves in two-dimensional $C_{\textrm{6v}}$ phoxonic crystals. The phoxonic crystals are created by arranging the silicon columns periodically in the air background. We propose a multi-objective optimization framework based on the NSGA-II collaborated with the finite element approach, where the bandgaps of acoustic and electromagnetic waves are treated separately as the objective values. The topological nature of bandgaps is determined by analyzing the positional relationships of paired degenerate modes through the modal field calculations, enabling the customization of one of the two bandgaps within the same unit cell. Unlike traditional approaches relying on the  band inversion to induce topological phase transitions, the proposed approach directly generates a pair of unit cells with distinct topological properties for both wave types, achieving the maximum bandgap matching in each case. We further demonstrate the existence of the  pseudospin-dependent topological edge states for both acoustic and electromagnetic waves, verifying their unidirectionality and robustness against backscattering and defects. This work establishes a systematic strategy  for customizing phoxonic topological states, offering a new avenue for the inverse design of multi-functional devices based on both sound and light.
\end{abstract}



\begin{keyword}
Phononic crystals, Photonic crystals, Topological  insulators, Inverse design, Topological optimization


\end{keyword}

\end{frontmatter}



\section{Introduction}
\label{sec1}

Photonic crystals (PtCs) \cite{joannopoulos1997photonic}, composed of materials with different dielectric constants arranged periodically, and phononic crystals (PnCs) \citep{kushwaha1993acoustic}, their acoustic counterparts made of materials with vastly different acoustic properties, both exhibit bandgaps that prohibit the transmission of waves within certain frequency ranges. Due to their geometrical similarity, it becomes possible to manipulate both phonons and photons within a periodically phononic (or photonic) structure simultaneously. For example, periodic silicon rods embedded in the air background can act as PtCs for electromagnetic waves and meanwhile can be considered as PnCs for airborne sounds. This concept, known as phoxonic crystals (PxCs), or optomechanical crystals, was introduced initially \citep{maldovan2006simultaneous} and has since been widely studied for their phoxonic (both photonic and phononic) bandgaps \citep{mohammadi2010simultaneous, ma2014investigation} and enhancing optomechanical interactions \citep{gavartin2011optomechanical, rolland2012acousto, el2013analysis}. 

By leveraging the unidirectional conductivity of topologically protected edge or boundary states and the robustness to geometrical imperfections, topological insulators (TIs) for classical waves are making significant strides \cite{ni2023topological}. As artificially periodic materials, phononic TIs (PnTIs) \cite{he2016acoustic,lu2016valley, zhang2018topological,tang2023topological} and photonic TIs (PtTIs) \citep{khanikaev2013photonic,lu2014topological, yang2018visualization} greatly enhance wave manipulation capabilities and have been extensively studied to explore their potentials in waveguiding, focusing and filtering, etc. Notably, one of the significant challenges is designing TIs that can simultaneously control two or even more types of classical waves. The realization of phoxonic TIs (PxTIs), which possess topological states of both acoustic and electromagnetic waves, is expected to enable robust transport of dual waves. Additionally, topological states achieved through optomechanics \citep{peano2015topological, ren2022topological} and other exotic phenomena such as phoxonic rainbow trapping \citep{ding2022simultaneous} and optomechanically induced transparency \citep{zangeneh2020topological} are gaining increasing attention.

Several efforts have been devoted to the studies of phoxonic topological edge states (TESs) \citep{ma2022topological, xia2019topologically, hu2022simultaneous, zhao2023topological, lei2022coexistence}. Ma \textit{et al.} \citep{ma2022topological} proposed one-dimensional (1D) PxTI analogs to the Su-Schrieffer-Heeger model based on PxC cavity chains. For two-dimensional (2D) PxTIs, Xia \textit{et al.} \citep*{xia2019topologically} demonstrated that honeycomb latticed PxCs exhibit both acoustic and electromagnetic TESs, with distinct mechanisms underlying the band degeneracy for different types of waves. Subsequently, dual phononic and photonic TESs based on the same mechanism, i.e., quantum spin Hall effect (QSHE) \citep{hu2022simultaneous} or  quantum valley Hall effect (QVHE) \citep{zhao2023topological}, were demonstrated. Lei \textit{et al.} \citep{lei2022coexistence} proposed a higher-order PxTI in a square lattice capable of localizing both acoustic and electromagnetic waves at a specific position, i.e., topological corner states. Such higher-order PxTIs beyond the traditional bulk-boundary correspondence could host lower-dimensional boundary states, such as topologically protected corner states in gapped energy bands of edge states, offering novel approaches for energy confinement and amplification \citep{xie2021higher}.

However, the aforementioned PxTIs were primarily designed through empirical methods or trial-and-error procedures, which restricts the ability to systematically customize topological states. Algorithms, including topology optimization, have been widely used for the inverse design of artificially periodic materials with unique wave properties such as bandgaps \citep{sigmund2003systematic,meng2017microstructural,dong2024inverse} or negative effective parameters \citep{meng2016topology,li2019topology}. 

Regarding PnTIs or PtTIs, several strategies \citep{chen2022inverse} have been proposed to optimize the topological bandgaps and achieve topological phase transitions. These strategies are essential for advancing from focusing on a single target to addressing dual targets. Christiansen \textit{et al.} \citep{christiansen2019topological,christiansen2019designing} used a black-box design strategy to optimize backscatter-protected unidirectional transport in the pre-assembled TIs system. A more common practice is to inversely design promising unit cells (UCs). It is feasible to choose a pair of UCs before and after the phase transition critical point, or more directly, to design two topologically matched single UCs. A metric  called frequency-averaged localized state density (LDOS) is widely adopted, and works well for customizing specific patterns of UCs. This metric is proportional to the power radiated by the time-harmonic source at a reasonably chosen location \citep{liang2013formulation,lin2016enhanced, chen2019topology,cai2023inverse, chen2021creating, zheng2023switchable,dong2021customizing}. However, this method is not reliable in obtaining topological bandgaps for complex structures due to the lack of evaluation of the overall energy bands. Another approach is to obtain the energy bands directly through an iterative computational process, resulting in more intuitive and accurate access to spectral features such as degeneracy or bandgap \citep{nanthakumar2019inverse,chen2022second, chen2023customizable, li2024topology}. Note that some additional steps are required to obtain the topological invariants or determine the topological properties for the calculated  bandgaps. However, solving topological invariants in real time during iteration is clearly impractical, as it consumes a lot of computational resources. There is a strategy to recognize the topological properties of the bandgaps and thus mode inversions by the modal assurance criterion (MAC) \citep{lu2021double, luo2021moving, du2020optimal}. In the present work, we will adopt a more ingenious approach to identify the positional relation of modes and then determine the topological properties.

The inverse designs for TESs in PnTIs / PtTIs are relatively well-established, and other investigations have also been conducted for three-dimensional (3D) TIs \citep{kim2023automated,luo2023efficient}, reconfigurable TIs \citep{zhuang2022inverse} and using tools such as machine learning \citep{long2019inverse,peano2021rapid, he2021inverse,li2023machine,du2023higher}, among others. Whereas, the inverse design of PxTIs for both acoustic and electromagnetic waves has not been extensively studied. To the best of our knowledge, only the second-order $C_{\textrm{4v}}$ phoxonic TESs in square lattices have been realized by topology optimization recently \citep{chen2024design}, while the UCs with opposite topological natures are obtained by different choices in the same lattice, similar to Ref. \citep{lei2022coexistence}. Comparatively, the QSHE-based inverse design of $C_{\textrm{6v}}$ PxC is of more interest because the unidirectional pseudospin dependent TES has stronger topological properties. The geometrical demands of the two topological bandgaps may be mutually exclusive, and we can only find compromise solutions to be compatible with them, which requires rational multi-objective optimization. It is also critical to match the bandgaps and topological properties of the paired QSHE-based UCs under acoustic and electromagnetic conditions, respectively. 

In this work, we present a multi-objective optimization (MOP) paradigm for  pseudospin-dependent acoustic and electromagnetic wave TESs in $C_{\textrm{6v}}$ PxCs. The structure of the paper is organized as follows: Section~\ref{sec2} describes the pixelated geometric characterization, the physical mechanism of TESs, and the optimization design process. In Section~\ref{sec3}, we provide detailed optimization formulas for the promising UCs and verify the TESs, including unidirectional transmission through numerical simulations, followed by the conclusions in Section~\ref{sec4}.

\begin{figure*}[h]
	\centering
	\includegraphics{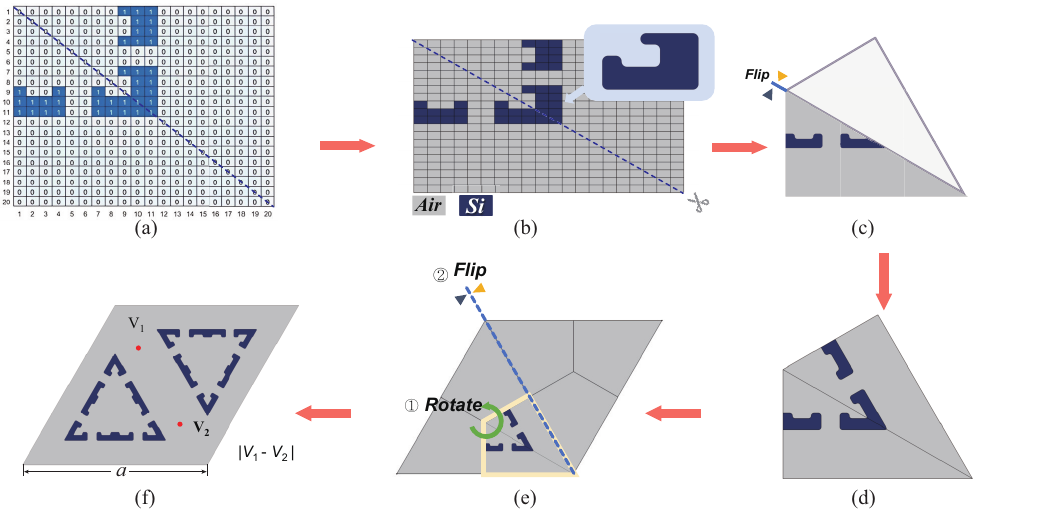}
	\caption{Procedure for constructing the PxC geometry using the 0/1 coding method. (a) shows a sample of 0/1 coded matrix after the graphical processing, where 0 and 1 represent air and silicon, respectively. The matrix is projected onto a rectangular array in (b),  and then flipped along the main diagonal in (c). The connected air domain and the merged silicon domains form a 1/6 UC geometry together in (d). Finally, after a rotation and flip process as shown in (e), the final PxC UC with the $C_{\textrm{6v}}$ symmetry is obtained in (f).}
	\label{fig1}
\end{figure*}

\section{Model and method}
\label{sec2}

\subsection{Geometry of UCs}
\label{subsec2}

To assemble the 2D PxC UCs in the hexagonal lattice, we use a 0/1 encoded matrix with the mirror symmetry along the main diagonal, as illustrated in  Fig.~\ref{fig1}(a). Herein, 0 represents the air background and 1 denotes the solid material (silicon). The encoded matrix is projected into a rectangular domain, where each pixel block is correspondingly changed into a small rectangle with the length-to-width ratio of \(\sqrt{3}\), as shown in Fig.~\ref{fig1}(b). However, for this discrete encoding method, it is difficult to generate a smooth geometry by limited pixels. To address this, the filleting operation is applied to the corners of merged solid blocks, smoothing the geometry at boundaries and hence preventing singular solutions. Additional graphical processing is used to remove weak connections and prevent checkerboard patterns. The lower-left triangle of the rectangle in Fig.~\ref{fig1}(b) is extracted and then flipped along the main diagonal, forming one-sixth of the PxC UC, as displayed in Figs. \ref{fig1}(c) and \ref{fig1}(d). This quadrilateral section is then rotated twice (by 60° and 120°) and subsequently mirrored along the mirror axis, as illustrated in Fig.~\ref{fig1}(e). The resulting PxC UC with the $C_{\textrm{6v}}$ symmetry is shown in Fig.~\ref{fig1}(f), where \(V_{1}\) and \(V_{2}\) indicate the field-value acquisition points for the subsequent modal type analysis.

\subsection{Governing equations }
The governing equation for acoustic waves propagating in the inhomogeneous and isotropic medium can be expressed as
\begin{equation}
	\nabla \cdot \left( \frac{1}{\rho \left( \mathbf{r} \right)}\nabla p\left( \mathbf{r},t \right) \right) 
	- \frac{1}{K\left( \mathbf{r} \right)}\frac{{{\partial }^{2}}p\left( \mathbf{r},t \right)}{\partial {{t}^{2}}} = 0,
\end{equation}
where \( p\left( \mathbf{r},t \right) \) is the sound pressure; \( \rho \left(\mathbf{r}\right) \) is the mass density; \( K\left( \mathbf{r}\right) \) is the bulk modulus; $\mathbf{r}$ and $t$ denote the position vector and the time, respectively.

On the other hand, the electromagnetic waves in 2D systems can be classified into two modes according to the polarization: the transverse magnetic (TM) mode (with the out-of-plane electric field component \(E_z\) and the in-plane magnetic field components $H_x$ and $H_y$), and the transverse electric (TE) mode (with the out-of-plane magnetic field component \(H_z\) and the in-plane electric field components $E_x$ and $E_y$). In this study, we focus on the electromagnetic TM mode, whose governing equation is given by
\begin{equation}
	 \nabla^{2} E_{z}\left( \mathbf{r},t \right) -\varepsilon \left(\mathbf{r}\right)\mu_{0}\frac{{{\partial }^{2}}E_{z}\left( \mathbf{r},t \right)}{\partial {{t}^{2}}}=0,
\end{equation}
where \( \varepsilon \left( \mathbf{r} \right) \) and $\mu_{0}$  represent the dielectric constant and permeability of vacuum, respectively.

\begin{figure}[H]
	\centering
	\includegraphics{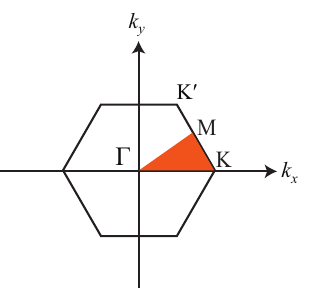}
	\caption{The first IBZ of the 2D hexagonal lattice and the high-symmetric points.}
	\label{fig2}
\end{figure}

The finite element method (FEM) enables the numerical solution of complex structures across various physical fields. In this study, we employ the commercial software COMSOL Multiphysics to conduct the numerical calculations. Due to the periodicity of 2D PxCs, the Bloch periodic boundary conditions are applied to the outer boundaries of PxC UCs. By solving the eigenvalue problem as the Bloch wave vector $\textbf{k}$ varies along the boundary of the first irreducible Brillouin zone (IBZ) (see Fig. \ref{fig2}), the band structures of both acoustic and electromagnetic waves can be obtained.

\subsection{Phoxonic pseudospin-dependent TESs}

For both acoustic and electromagnetic waves, the formation of a double Dirac cone in the hexagonal lattice could be achieved through band folding \citep{wu2015scheme} or accidental degeneracy \citep{sakoda2012double, he2016acoustic}. This feature is crucial for the realization of pseudospin-orbit coupling in these bosonic systems. Therefore, in the subsequent discussion we will apply such a feature to both wave types for the generation of pseudospin-dependent TESs. The $C_{\textrm{6v}}$ symmetry group includes two 2D irreducible representations, which leads to paired dipole (\( p_x \), \( p_y \)) and quadrupole (\( d_{xy} \), \( d_{x^2 - y^2} \)) modes at point $\Gamma$. Empirically, these \textit{p}-type and \textit{d}-type modes consistently appear within the first to seventh bands. For topologically trivial crystals, the corresponding bandgaps form with the \textit{d}-type modes positioned above the \textit{p}-type pairs. In contrast, for topologically nontrivial bandgaps, the \textit{p}-type modes are found above the \textit{d}-type modes. For instance, if a PnC is composed of a hexagonal lattice of solid columns in the air background, the variation of column radius leads to the PnC transforming from a trivial into a nontrivial phase. At the critical radius of \( 0.3928a \) (where \( a \) is the lattice constant), the bandgap closes, resulting in a four-fold accidental degeneracy \citep{he2016acoustic}. A combination of crystals with distinct topological characteristics can give rise to pseudospin-dependent TESs at the interface.

To achieve TESs for both acoustic and electromagnetic waves, we employ the topology optimization approach to inversely design two PxC UCs with distinct topological properties, referred to as the first UC (FUC) and the counterpart UC (CUC), rather than selecting them before and after the Dirac point. Notably, customizing the topological properties of the FUC and CUC in advance is crucial to ensure the topological phase transitions of acoustic and electromagnetic waves simultaneously. We adopt a strategy to compute the difference of eigenfields at two specific points $|V_1 - V_2|$, as depicted in Fig.~\ref{fig1}(e). Since these data collection points align with the modal distribution of \( p_x \) modes, enabling us to extract the maximum value from a set of four modes (\( p_x \), \( p_y \), \( d_{xy} \) and \( d_{x^2 - y^2} \)). This allows us to identify the \( p_x \) modes and subsequently determine the topological properties based on the spatial relationship between the pairs of the \textit{p}-type  and \textit{d}-type modes.

\subsection{NSGA-II}
\begin{figure}[h]
	\centering
	\includegraphics{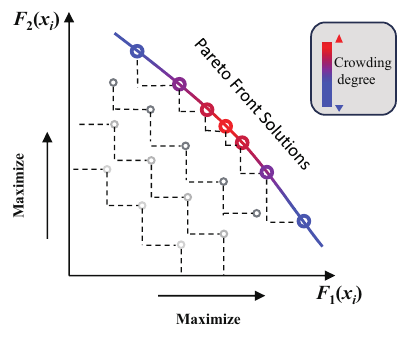}
	\caption{Schematic illustration of the optimization processes for the NSGA-II and its Pareto front.}
	\label{fig3}
\end{figure}

In this work, we utilize the multi-objective optimization strategy (NSGA-II) \cite{Deb2002A} to simultaneously maximize both the objective values for acoustic and electromagnetic waves, aiming to achieve 2D PxTIs.  The NSGA-II excels at maintaining solution diversity and balancing conflicting objectives, which makes it particularly suited for multi-objective optimization tasks. To date, some studies have applied this method to the optimization design of periodically artificial materials that possess particular wave functions, such as lightweight metamaterials with ultra-wide bandgaps \cite{yan2023multi}, and acoustic black holes \cite{bao2024ultra}.

\begin{figure}[h]	
	\centering
	\includegraphics{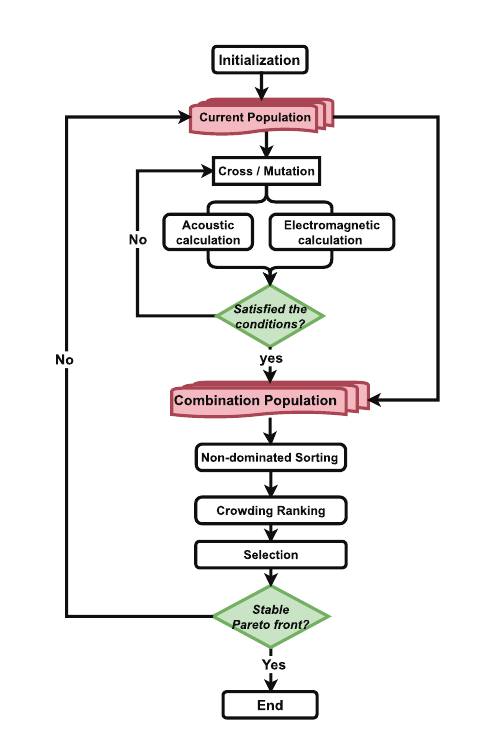}
	\caption{Flowchart of the optimization process for PxTIs based on the NSGA-II, where the loop is stopped and the optimal solution set is obtained after a stable Pareto front is reached.}
	\label{fig4}	
\end{figure}
\vspace{1cm}

The core of the NSGA-II involves the non-dominated sorting of the entire population to establish dominance hierarchies. The dominance means one individual is superior to another in at least one objective. Figure~\ref{fig3} provides a schematic of the Pareto front, where the horizontal and vertical axes are the two objective functions to be maximized simultaneously. The solutions are classified into multiple fronts based on the dominance relationships, with the highest level being the Pareto front, which is not dominated by any other individual. The crowding distance (depicted by the colors in Fig.~\ref{fig3}) indicates the density of individuals in the current front. During the selection process, the individuals with lower crowding distance are prioritized, ensuring the diversity when selecting a finite number of solutions from the combined population.

The basic optimization process for the 2D PxC UCs is shown in Fig.~\ref{fig4}. It starts with the population initialization, followed by generating new individuals through the crossover and mutation operations. The eigenfrequencies for acoustic and electromagnetic modes are calculated using FEM for each individual to obtain the target values, which are evaluated based on the frequencies and topological properties of bandgaps. The individuals meeting the criteria are added to the combination population. The non-dominated sorting and the crowding distance calculations are then performed to maintain the diversity. A subset of the combined population is selected to form the current population, and the process repeats until the convergence at the stable Pareto front. For detailed NSGA-II parameters used in this work, please refer to \ref{app1}.

\section{Results and discussion}
\label{sec3}
In this section, the topological optimization formulas for the pairs of PxC UCs (i.e., the FUC and CUC) are firstly presented, and then some PxC examples are given to demonstrate the ability of our algorithm to customize the frequencies and topological properties of a bandgap. At last, one-way and robust pseudospin-dependent TESs for both acoustic and electromagnetic waves are verified.

\subsection{Optimization of PxC FUCs}
Equation~(\ref{eq3}) presents the MOP fitness function for the PxC FUCs. Variables \textit{$x_{l}$} are the pixels of $n_{x} \times n_{y}$ coding matrix for the UC geometry shown in Fig.~\ref{fig1}. In the MOP framework, the two fitness parameters, \textit{$W_{a}$} and \textit{$W_{e}$}, denote the normalized bandgap widths of acoustic and electromagnetic waves, respectively. Equation~(\ref{eq4}) provides the formula for the normalized gap width \textit{$W_{\theta}$}, where \(f_{\theta}^{\text{c}}\) represents the center frequency, and $G_{\theta}^{\text{up}}$ and \(G_{\theta}^{\text{low}}\) are the upper and lower boundaries of the \(\theta\)-type bandgap, respectively. The subscript \(\theta\) indicates the type of waves, where \(\textit{a}\) stands for acoustic waves and \(\textit{e}\) for electromagnetic waves (TM mode). In general, empirically designed PxCs typically struggle to achieve the desired bandgap frequencies, especially the realization of topological bandgaps for two wave types within a single structure. 
\\
\\
\noindent\textbf{Maximize}: MOP(\textit{$x_{l}$}), ${{x}_{l}}=0/1\ \ (l\text{ }=1,\ 2,\,\ldots \ n_{x}\times n_{y})$

\begin{flalign}
	&\text{MOP}({{x}_{l}}) = \begin{cases}\label{eq3}
		({{W}_{a}},\ {{W}_{e}}) & \text{if }\left| f_{a}^{\text{c}} - f_{a}^{\text{ref}} \right| \leq 0.05 {{W}_{\text{a}}} \\
		\left( -\left| f_{a}^{\text{c}} - f_{a}^{\text{ref}} \right|, {{W}_{e}} \right) & \text{if }\left| f_{a}^{\text{c}} - f_{a}^{\text{ref}} \right| > 0.05 {{W}_{\text{a}}}
	\end{cases} &\\	
	&\textbf{where:} &\notag\\
	&{{W}_{\theta}} = \frac{G_{\theta}^{\text{up}} - G_{\theta}^{\text{low}}}{f_{\theta}^{\text{c}}},\ \ f_{\theta}^{\text{c}} = \frac{G_{\theta}^{\text{up}} + G_{\theta}^{\text{low}}}{2}, \ \theta = a, e &\label{eq4}
\end{flalign}
\noindent\textbf{Subject to:}
\begin{flalign}
	&{{f}_{\theta, 1}}(\Gamma, x_{l}) = {{f}_{\theta, 2}}(\Gamma, x_{l}), \quad{{f}_{\theta, 3}}(\Gamma, x_{l}) = {{f}_{\theta, 4}}(\Gamma, x_{l}) &\label{eq5} \\
	&G_{\theta}^{\text{up}} > G_{\theta}^{\text{low}} &\label{eq6}\\
	&G_{\theta}^{\text{up}} = \min \{ f_{\theta, 3}(k, x_{l}) \}, \quad \forall k \in \text{IBZ}&\label{eq7}\\
	&G_{\theta}^{\text{low}} = \max \{ f_{\theta, 2}(k, x_{l}) \}, \quad \forall k \in \text{IBZ} &\label{eq8}\\	
	&\begin{cases} \label{eq9}\{ f_{\theta, (1,2)} \} = \{ f_{\theta, (p_1,p_2)} \} \land \{ f_{\theta, (3,4)} \} = \{ f_{\theta, (d_1,d_2)} \} \\
	\{ f_{\theta,\ (1,2)} \} = \{ f_{\theta, (d_1,d_2)} \} \land \{ f_{\theta, (3,4)} \} = \{ f_{\theta, (p_1,p_2)} \} 
	\end{cases} &
\end{flalign}

\begin{figure}[h]
	\centering
	\includegraphics{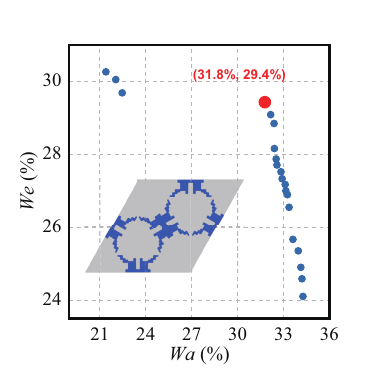}
	\caption{FUC geometry with the customized acoustic frequency of 1.1, which is selected manually by the Pareto front. The horizontal and vertical axes represent the normalized acoustic and electromagnetic banggap widths, respectively.}
	\label{fig5}
\end{figure}
\vspace{1cm}

It is well known that the geometrical configuration of PxC UCs can influence the band diagrams of acoustic and electromagnetic waves simultaneously. As a result, predicting the acoustic bandgap solely from the electromagnetic bandgap, and vice versa, is often challenging. If the bandgap frequencies for both wave types are strictly constrained, a solution may not inherently exist. In this work, the optimization algorithm allows for the predefined placement of one bandgap, such as the acoustic bandgap, by setting a reference frequency, \( f_{a}^{\text{ref}} \) in advance. As described in Equation~(\ref{eq3}), if the distance between the reference frequency \( f_{a}^{\text{ref}} \) and the normalized center frequency \( f_{a}^{c} \) of the acoustic bandgap exceeds a threshold (here $0.05 W_{a}$), the negative deviation is used as the acoustic target fitness \textit{$W_{a}$}. Equations~(\ref{eq5}-\ref{eq9}) provide the constraints and preconditions for both acoustic and electromagnetic waves. Equation~(\ref{eq5}) represents four energy bands from the bottom up, consisting of paired $p$-type or $d$-type modes degenerate at point $\Gamma$, without the interference from other energy bands. Equation~(\ref{eq6}) ensures that a bandgap opens between the $p$-type and $d$-type bands. Moreover, it is necessary to perform a sufficient $\textbf{k}$-space scanning to accurately capture the extremes of the upper [Equation~(\ref{eq7})] and lower [Equation~(\ref{eq8})] bands of the target bandgap. As described by Equation~(\ref{eq9}), the proposed formulation supports specifying the topological phases of the target FUC for either the acoustic or electromagnetic mode, through the modal field value evaluation method.

In this work, we consistently use the normalized frequencies, defined as ${fa/c_{\mathrm{air}}}$ and ${fa/c_{0}}$ for acoustic and electromagnetic waves, respectively, where $f$ is the frequency, $a$ is the lattice constant, $c_{\text{air}}$ is the sound speed in the air, and $c_{\text{0}}$ is the light speed in the vacuum. The solid material used  in this work is silicon, with the material parameters as follows: the mass density $\rho_{\mathrm{si}} = 2331$ $\text{kg/m}^3$, the velocity of longitudinal waves $c_{\mathrm{si}} = 8950$ $\text{m/s}$ and the refractive index $n_{\mathrm{si}} = \sqrt{13}$. Additionally, the mass density, sound speed and refractive index of the air are  $\rho_{\mathrm{air}} = 1.204$ $\text{kg/m}^3$, $c_{\mathrm{air}} = 343$ $\text{m/s}$ and $n_{\mathrm{air}} = 1$, respectively.

We select an individual with the acoustic reference frequency of \( f_{a}^{\text{ref}}=1.1\), characterized by a trivial acoustic bandgap and a nontrivial electromagnetic bandgap, as the example for the subsequent discussion. Figure~\ref{fig5} gives the Pareto front after the optimization processes. One of the individuals is manually selected, as illustrated in the inset. It is seen that the ring-shaped solids at the lattice points are approximately connected to each other, and the centered cavity is in conduction with the outside air domain. For the optimization results for PxC FUCs with other frequencies and topological properties, please refer to \ref{app2}.

\begin{figure}[h]
	\centering
	\includegraphics{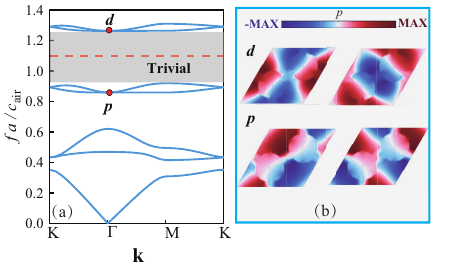}
	\caption{(a) Acoustic band structure with a topologically trivial bandgap and (b) the corresponding field profiles of the \textit{d}-type and \textit{p}-type modes for the PxC FUC with the customized normalized frequency of 1.1.}
	\label{fig6}
\end{figure}

\begin{figure}[h]
	\centering
    \includegraphics{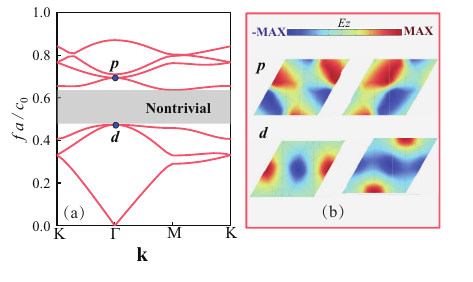}
	\caption{(a) TM mode band structures with a topologically nontrivial bandgap and (b) the corresponding field profiles of the \textit{p}-type and \textit{d}-type modes (b) for the PxC FUC.} 
	 \label{fig7}
\end{figure}

Figure~\ref{fig6}(a) shows the acoustic band structure of this PxC FUC while its modal profiles for the \textit{p}-type and \textit{d}-type modes at point $\Gamma$ are shown in Fig.~\ref{fig6}(b). Note that the topological property can be judged by the order of \textit{p}-type and \textit{d}-type paired modes. Here, the \textit{d}-type modes are above the \textit{p}-type modes and hence the acoustic bandgap is topologically trivial. Besides, the band structure of this individual shows a better match between its bandgap-center-frequency and the reference frequency preset in advance, demonstrating the algorithm's ability to customize the location of the acoustic bandgap. The results for the electromagnetic TM mode are shown in Fig.~\ref{fig7}, in which the \textit{p}-type modes are above the \textit{d}-type modes, by same strategy, so that the electromagnetic bandgap is topologically nontrivial. For the designed PxC UC, the normalized acoustic and electromagnetic bandgap widths are 31.8 $\%$ and  29.4$\%$, respectively. We find that different topological properties for acoustic and electromagnetic waves could avoid geometrical competition and support compatibility with wide bandgaps for dual waves. This is conducive to match the PxC CUCs with overlapping bandgaps that possess distinct topological properties in the subsequent optimization process.

\subsection{Optimization of PxC CUCs}

\begin{figure*}[h]	
	\centering	
	\begin{flalign}
	    &\textbf{Maximize}:\text{MOP}(\textit{$x_{l}$}) = [fit_{a}, fit_{e}]^{\mathrm{T}}, \quad {{x}_{l}}=0/1\ \ (l=1, 2, \ldots, nx \times ny) \notag & \\
		&{{fit}_{\theta }} = \frac{{{L}_{\theta }}}{G_{\theta }^{\text{up}}-G_{\theta }^{\text{low}}} - \frac{{{g}_{\theta }}}{(G_{\theta }^{\text{up}}-G_{\theta }^{\text{low}})+(G_{\theta }^{\text{up}*}-G_{\theta }^{\text{low}*})} , \quad \theta = a, e & \label{eq10} \\
		&\begin{cases} \label{eq11}
			{{g}_{\theta }}=\text{max}\left( G_{\theta }^{\text{up}*}-G_{\theta }^{\text{low}},G_{\theta }^{\text{up}}-G_{\theta }^{\text{low}*} \right) \\ 
			{{L}_{\theta }}=0 
		\end{cases} \qquad  \text{if }[G_{\theta }^{\text{low}}, G_{\theta }^{\text{up}}] \cap [G_{\theta }^{\text{low}*}, G_{\theta }^{\text{up}*}] = 0 & \\
		&\begin{cases} \label{eq12}
			{{g}_{\theta }}=0 \\ 
			{{L}_{\theta }}= \min(G_{\theta }^{\text{up}}, G_{\theta }^{\text{up}*}) - \max(G_{\theta }^{\text{low}}, G_{\theta }^{\text{low}*}) 
		\end{cases} \quad \text{if }[G_{\theta }^{\text{low}}, G_{\theta }^{\text{up}}] \cap [G_{\theta }^{\text{low}*}, G_{\theta }^{\text{up}*}] \neq 0 &\\
		&{{f}_{\theta \text{, 1}}}\text{(}\Gamma\text{, }{{x}_{\text{e}}}\text{) = }{{f}_{\theta \text{, 2}}}\text{(}\Gamma\text{, }{{x}_{\text{e}}}\text{) = }{{f}_{\theta \text{, 3}}}\text{(}\Gamma\text{, }{{x}_{\text{e}}}\text{) = }{{f}_{\theta \text{, 4}}}\text{(}\Gamma\text{, }{{x}_{\text{e}}}\text{)}& \label{eq13}\\
		&G_{\theta }^{\text{up}} > G_{\theta }^{\text{low}}& \label{eq14}\\
		&G_{\theta }^{\text{up}} = \min \left\{ {{f}_{\theta \text{, 3}}}\text{(}k,{{x}_{\text{e}}}\text{)} \right\}\text{, }\forall k\in \text{IBZ}& \label{eq15} \\
		&G_{\theta }^{\text{low}} = \max \left\{ {{f}_{\theta \text{, 2}}}\text{(}k,{{x}_{\text{e}}}\text{)} \right\}\text{, }\forall k\in \text{IBZ}& \label{eq16}\\		
		&\begin{cases} \label{eq17} \{ f_{\theta, (1,2)} = f_{\theta, (p_1,p_2)}\} \land \{ f_{\theta, (3,4)} = f_{\theta, (d_1,d_2)} \} \quad \text{if }\{ f_{\theta, (1,2)}^{*} = f_{\theta, (d_1,d_2)}^{*} \} \land \{ f_{\theta, (3,4)}^{*} = f_{\theta, (p_1,p_2)}^{*} \} & \\
			\{ f_{\theta, (1,2)} = f_{\theta, (d_1,d_2)} \} \land \{ f_{\theta, (3,4)} = f_{\theta, (p_1,p_2)} \} \quad \text{if }\{ f_{\theta, (1,2)}^{*} = f_{\theta, (p_1,p_2)}^{*} \} \land \{ f_{\theta, (3,4)}^{*} = f_{\theta, (d_1,d_2)}^{*} \}
		\end{cases} &
	\end{flalign}
\end{figure*}

In this subsection, we will then present the optimization formulas for the PxC CUCs, which share some similarities in interpretation with those for the FUC. For a PxC CUC, the goal of the CUC is to optimize the design to maximize bandgap overlap of the FUC in Fig.~\ref{fig5}. For the $\theta$-type waves, the nonlinear expression $fit_\theta$ [see Equation~(\ref{eq10})] describes the dynamic relationship between the target bandgap \([G_{\theta }^{\text{low}}, G_{\theta }^{\text{up}}]\) of the CUC and the established  bandgap \([G_{\theta }^{\text{low}*}, G_{\theta }^{\text{up}*}]\) extracted from the FUC in the previous stage. The subscript $\theta$ is \textit{a} for sound waves and \textit{e} for electromagnetic waves. As described in Equation~(\ref{eq11}), the width of overlapped bandgaps $L_{\theta}$ is set to 0 if the two bandgaps do not intersect. Meanwhile, the parameter $g_{\theta}$, which denotes the distance between \([G_{\theta }^{\text{low}}, G_{\theta }^{\text{up}}]\) and \([G_{\theta }^{\text{low}*}, G_{\theta }^{\text{up}*}]\), is also evaluated. If the bandgaps of the FUC and CUC overlap, $L_{\theta}$ is calculated according to Equation~(\ref{eq12}) and $g_{\theta}$ is set to 0, and then $fit_\theta$ is the ratio between the width of the overlapped bandgaps and the bandgap width of the CUC. Note that  Equations~(\ref{eq13}-\ref{eq16}) are the same as those in the formulas for the FUC, ensuring that the target bandgaps are generated based on the QSHE approach. Equation~(\ref{eq17}) presents how the topological properties of the PxC CUC are determined by those of the FUC.

\begin{figure}[H]
	\centering
	\includegraphics{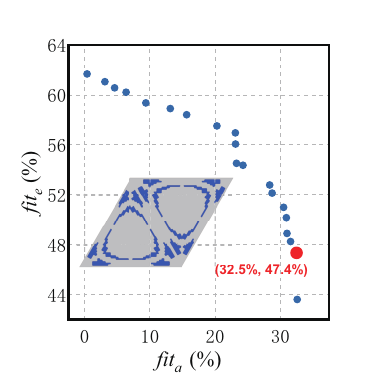}
	\caption{CUC geometry corresponding to the FUC with customized acoustic frequency of 1.1, which is selected manually by the Pareto front. The horizontal and vertical axes represent the degree of overlap between the two UCs in terms of acoustic and TM bandgaps, respectively.}
	\label{fig8}
\end{figure}

\begin{figure}[h]
	\centering
	\includegraphics{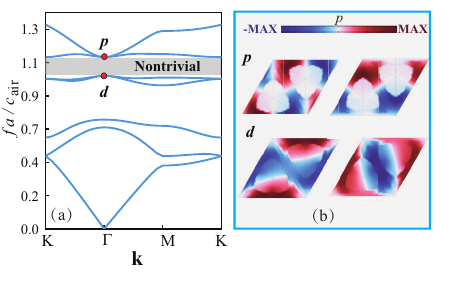}
	\caption{(a) Acoustic band structures with a topologically nontrivial bandgap and (b) the corresponding field profiles of the \textit{p}-type and \textit{d}-type modes for the PxC CUC.}
	\label{fig9}
\end{figure}

The optimization results for the PxC CUC, corresponding to the designed FUC [see Fig. \ref{fig5}(b)] at the acoustic reference frequency of 1.1, are shown in Fig.~\ref{fig8}. The Pareto front is displayed, where $fit_{a}$ and $fit_{e}$ are represented by the horizontal and vertical axes, respectively. These parameters relate to the degree of overlap between the acoustic (or electromagnetic) bandgaps of the CUC and FUC. To maximize the overlap of  bandgaps for dual waves, we selected a specific individual (indicated by the red dot), whose geometry is shown in the inset of Fig. \ref{fig8}. A notable difference between the geometries of the CUC and FUC is evident. The matching rate is 32.5\% for the acoustic bandgap and 47.4\% for the electromagnetic bandgap. 

\begin{figure}[h]
	\centering
    \includegraphics{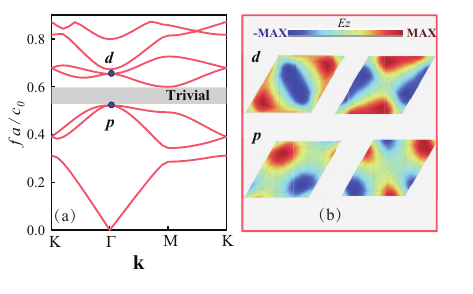}
	\caption{(a) TM mode band structures with a topologically trivial bandgap and (b) the corresponding \textit{d}-type and \textit{p}-type modes for the PxC CUC.}
	 \label{fig10}
\end{figure}

The acoustic band structure of the PxC CUC is plotted in Fig.~\ref{fig9}. Its mode distribution shows that the \textit{p}-type modes are positioned above the \textit{d}-type ones, signifying a nontrivial topology. This is contrary to the topological nature of the acoustic bandgap of the FUC [see Fig. \ref{fig6}]. Similarly, the band structure and the eigenmodes of electromagnetic waves [see Fig. \ref{fig10}] reveals that the electromagnetic bandgap is topologically nontrivial, that is, the electromagnetic bandgaps of the designed CUC and FUC possess distinct topological characteristics. Thus, one can infer that by combining these PxCs with different topological properties, acoustic and electromagnetic TESs could be achieved along the crystal interface simultaneously.

\subsection{Unidirectional phoxonic TESs}

\begin{figure}[h]
	\centering
	\includegraphics{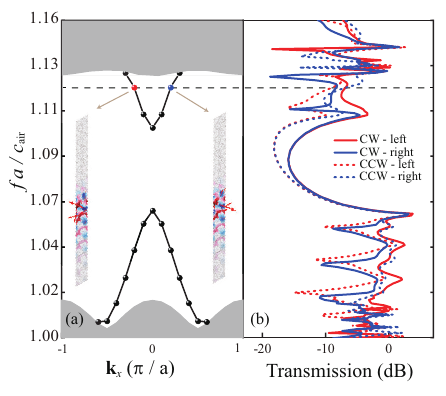}
	\caption{(a) Gapped acoustic TESs within the bulk bandgap. The insets show the eigenmodes of the TESs featuring opposite energy flows. (b) Transmission curves of acoustic waves under the CW and CCW rotational sources.}
	 \label{fig11}
\end{figure}

\begin{figure}[h]		
	\centering
	\includegraphics{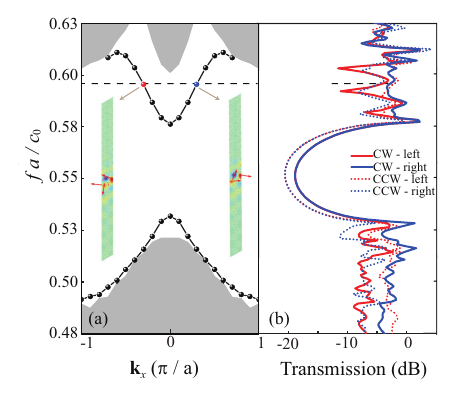}	
	\caption{(a) Gapped electromagnetic TESs within the bulk bandgap. The inset show the eigenmodes featuring opposite energy flows. (b) Transmission curves of electromagnetic waves under the CW and CCW rotational sources. }
	\label{fig12}
\end{figure}

Based on the optimized topological designs of the FUC (see Fig. \ref{fig5}) and CUC (see Fig. \ref{fig8}), we construct such a pair of PxC UCs with maximally matched bandgaps under both the acoustic and electromagnetic conditions, exhibiting distinct topological properties for dual waves. The ribbon super-cell, derived from a linear combination of two types of UCs (8+8) [see the inset of Fig. \ref{fig11}(a)], is employed to verify the emergence of acoustic and electromagnetic TESs. As shown in Fig.~\ref{fig11}(a), the acoustic TESs manifest in the bandgaps opened by the bulk energy bands (gray area). Nevertheless, due to geometrical differences between these two types of PxC UCs, the resulting symmetry breaking of the overall \(C_{6v}\) lattice leads to the edge states not bridging the bandgap. Meanwhile, the edge states exhibit distinct energy flow directions when the wave vectors are oppositely directed, demonstrating the  pseudospin-locking effect, as depicted in the insets of Fig.~\ref{fig11}(a). It is necessary  to note that this overlapping frequency range may not include the initial reference frequency \(f_{a}^{\mathrm{ref}}=1.1\), as our focus is the achievement of the maximum bandgap alignments between the two PxC UCs, for both acoustic and electromagnetic waves. On the other hand, the energy bands corresponding to the electromagnetic TESs emerge within the bulk bandgap, similar to the acoustic wave case, but with the opposite direction of chiral energy flow, as plotted in Fig. \ref{fig12}(a).

\begin{figure}[h]
	\centering
	\includegraphics{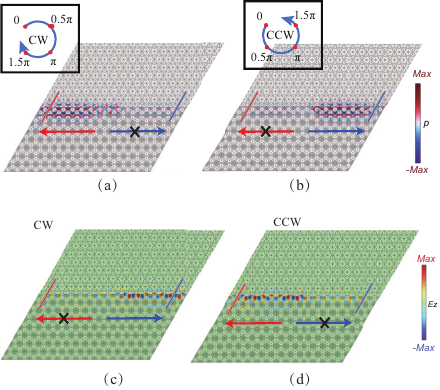} 
	\caption{Field distributions of the PxC structure with a straight interface between the inverse designed FUCs and CUCs: (a) leftward acoustic transmission (CW source), (b) rightward acoustic transmission (CCW source), (c) rightward electromagnetic transmission (CW source), and (d) leftward transmission (CCW source).
	}
	\label{fig13}
\end{figure}

To further validate the pseudospin-dependent unidirectional transmission of dual waves, we conduct the analysis of frequency responses for a finite-sized PxC structure containing a straight interface, as shown in Fig.~\ref{fig13}. To excite the pseudospin-dependent TESs, we utilize a rotating source composed of four precisely aligned points with a phase difference of \(0.5 \pi\), as illustrated in the insets of Fig.~\ref{fig13}. By alternating the phase positions between \(0.5 \pi\) and \(1.5 \pi\), both the clockwise (CW) and counterclockwise (CCW) rotational sources are generated. This source is positioned at the center of the structural interface, and the sound pressure is measured at each end of the interface. The transmission spectra of acoustic waves are displayed in Fig. \ref{fig11}(b). For the CW acoustic source, the transmission coefficient at the left end (red solid line) is lower than that at the right end (blue solid line) within the frequency ranges of TES bands, indicating the leftward transmission of acoustic waves. In contrast, the CCW acoustic source results in the rightward sound wave transmission, in which the pressure field at the right end (blue dashed line) is smaller than that at the left end (red dashed line). Additionally, the transmission spectra of electromagnetic waves are shown in Fig.~\ref{fig12}(b), and the distributions of out-of-plane electric fields at the TIs frequency marked in Fig. \ref{fig12} are shown in Figs. \ref{fig13}(c) and \ref{fig13}(d). The unidirectional propagation of electromagnetic waves is also observed, which is similar to the acoustic case. The response analysis demonstrates that the designed PxCs exhibit pseudospin-locked unidirectional TESs for acoustic and electromagnetic waves simultaneously.

\vspace{1cm}

\subsection{Robustness of phoxonic TESs}

\begin{figure}[h]
	\centering
	\includegraphics{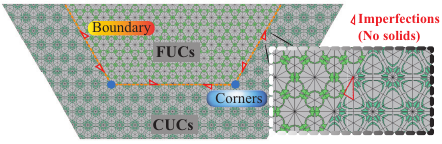}
	\caption{Structure used to manifest the the edge-bulk corresponding boundary states and higher-order corner states, where the point source are located in the middle of the edge with CW or CCW rotational sources.}
	\label{fig14}
\end{figure}	

\begin{figure}[htbp]
	\centering
	\includegraphics{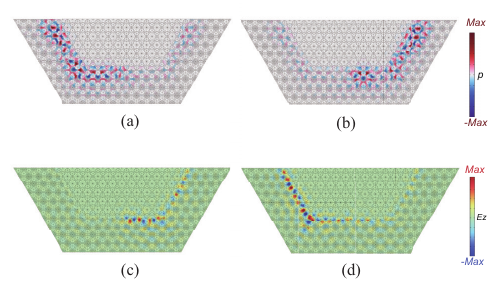}
	\caption{Field distributions of the bowl-shaped PxC structure with artificially introduced geometrical defects in the interface with FUCs and CUCs: (a) leftward acoustic transmission (CW source), (b) rightward acoustic transmission (CCW source), (c) rightward electromagnetic transmission (CW source), and (d) leftward  electromagnetic waves transmission(CCW source).}
	 \label{fig15}
\end{figure}

To verify the robustness of the unidirectional phoxonic TESs designed via the topology optimization, we assemble a bowl-shaped PxC structure and calculate its frequency responses for both acoustic and electromagnetic waves, as illustrated in Fig.~\ref{fig14}. In such a structure, the inner part consists of the FUCs (topologically trivial), which is surrounded by several layers of the CUCs (topologically nontrivial). The geometries of the PxC FUCs and CUCs are the same as those shown in Figs.~\ref{fig5}~and~ \ref{fig8}. Furthermore, we introduce some imperfections into the waveguide routine of the bowl-shaped structure, as illustrated in Fig.~\ref{fig14}. Herein, the solids (silicon) in several small triangles (indicated by the red boxes) are removed. We also apply the rotating point source (see Fig. \ref{fig13}) at the center of the interface of the defected  PxC structure. Figures \ref{fig15}(a) and \ref{fig15}(b) show the sound propagation under the CW and CCW  sources, respectively, while Figs. \ref{fig15}(c) and \ref{fig15}(d) show the results of electromagnetic waves. For either the acoustic or electromagnetic mode, these defects do not apparently affect the wave propagation. The results confirm that the topological PxC waveguide is insensitive to fabrication defects or impurities. Moreover, since the proposed bowl-shaped structure has corners in the waveguide path, it is further verified that there is an immunization effect against backscattering at the corners.

\begin{figure*}[!ht]		
	\centering
	\includegraphics{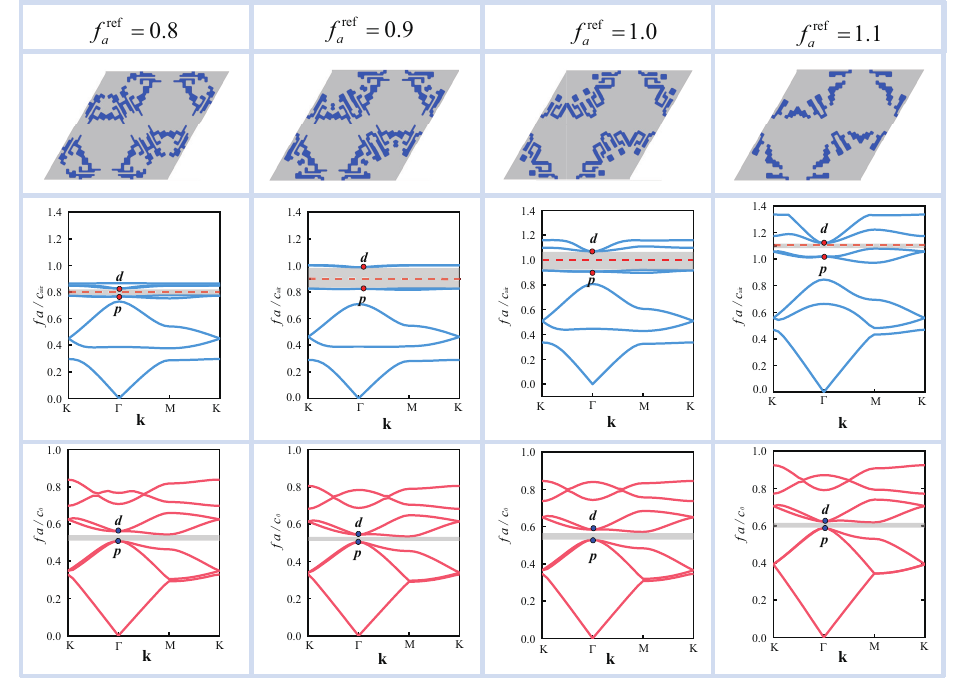}	
	\caption{Results of the optimized PxC UCs at the acoustic reference frequencies of 0.8, 0.9, 1.0, and 1.1, respectively, selected from the corresponding set of optimal solutions. The constraint is that both the acoustic and electromagnetic bandgaps have topologically trivial properties.}
	\label{fig16}
\end{figure*}

\begin{figure*}[!ht]
	\centering
	\includegraphics{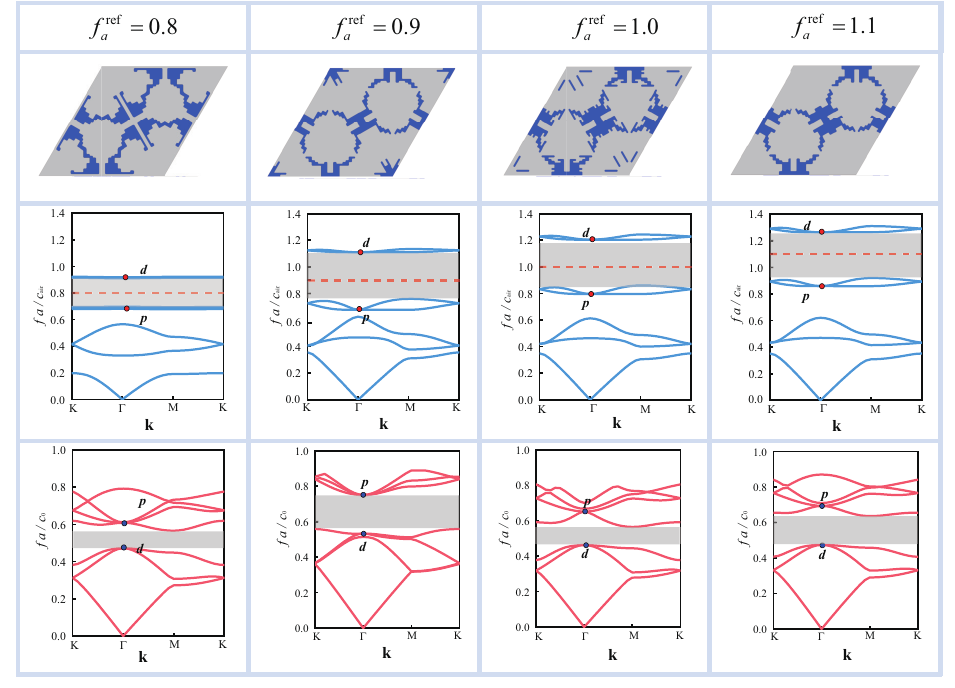}
	\caption{Results on the optimized PxC UCs at the acoustic reference frequencies of 0.8, 0.9, 1.0, and 1.1, respectively, selected from the corresponding set of optimal solutions. The constraint is that acoustic and electromagnetic bandgaps have topologically trivial and nontrivial properties, respectively.}
	\label{fig17}
\end{figure*}






\section{Conclusions}
\label{sec4}

In conclusion, we present a systematic inverse design paradigm for achieving the  pseudospin-dependent unidirectional TESs of acoustic and electromagnetic TM modes in 2D PxCs. The acoustic and electromagnetic waves are treated as distinct objective values in the multi-objective optimization algorithm NSGA-II. The feature that  the \( p \)-type and \( d \)-type modes appearing in pairs respectively is used to identify the target bandgaps. To accurately determine the topological properties for target bandgaps, we employ the method that identifies the order of paired \( p \)-type and \( d \)-type modes by leveraging the energy distribution characteristics of the \( p_x \) modes. The optimization results reveal that assigning different topological properties to the bandgaps of the two wave types  within a single UC leads to broader relative bandgap widths compared to assigning identical topological properties. The optimization algorithm enables the effective customization of bandgap frequencies for any one of the two wave types, exemplified here with the acoustic case. Through the two-step optimization process, we design a pair of PxC UCs exhibiting distinct topological properties as well as ensuring the optimal bandgaps matching for two types of waves. The 2D  PxTIs are  formed by combining these optimized UCs, demonstrating the unidirectional pseudospin-dependent TESs for both acoustic and electromagnetic waves. The full wave simulations validate the robustness of the resulting unidirectional TESs against geometrical defects and backscattering at structural corners. This work provides an inverse design formulation for the simultaneous realization of acoustic and electromagnetic topological states and can be used as a basis for enhancing the optomechanical interaction in PxC structures.

\section*{CRediT authorship contribution statement}
\textbf{Gang-Gang Xu}: Writing – original draft; Investigation; Methodology; Formal analysis; Validation; Conceptualization. \textbf{Xiao-Shuang Li}: Writing – review \& editing; Software; Methodology. \textbf{Tian-Xue Ma}: Conceptualization; Supervision; Writing – review \& editing; Funding acquisition. \textbf{Xi-Xuan Liu}: Formal analysis; Modeling. \textbf{Xiao-Wei Sun}: Writing – review \& editing. \textbf{Yue-Sheng Wang}: Funding acquisition; Supervision.

\section*{Declaration of Competing Interest}
The authors declare that they have no known competing financial interests or personal relationships that could have appeared to influence the work reported in this paper.

\section*{Data availability}
The data that support the findings of this study are available from the corresponding author upon reasonable request.

\section*{Acknowledgments}
This work is supported by National Natural Science Foundation of China (Grant Nos. 12372087, 12021002, 11991031, 12462011).

\appendix
\section{Parameters of NSGA-II}
\label{app1}
The pixel density is set to 20 in both \( n_x \) and \( n_y \), as shown in Fig. \ref{fig1}. Besides, the size of chamfers is 0.5\textit{d}, where \textit{d} denotes the width of a single pixel. To ensure the connectivity within the air region while keeping the solid regions separated, only one interconnected air domain (where the pixel value is 0) is maintained. The population comprises 20 individuals, with a crossover rate of 0.7 and a mutation rate of 0.1. Following the crossover operation, 12 new individuals are produced, and additional 4 individuals are generated through the mutation, yielding a combined population of 36. From this pool, 20 individuals are selected for the next generation based on the non-dominated sorting and the crowding distance. This process is repeated until a stable Pareto front is achieved, meaning no superior individual emerges.

Evolutionary algorithms, such as genetic algorithms, are generally more computationally intensive than sensitivity-based topology optimization methods. To improve efficiency, the parallel computing with 8 threads is utilized, reducing the computation time to around 1 minute per generation. The optimization process typically completes within tens to hundreds of generations.

\section{Examples FUCs}
\label{app2}
Here are additional examples of the optimized PxC FUCs, featuring normalized reference frequencies for the acoustic bandgap at 0.8, 0.9, 1.0, and 1.1. These geometries are selected from the optimal solution set and exhibit strong performance for both objectives. By applying the constraint condition in Equation~\ref{eq9}, we filter out the individuals that fail to meet the criteria based on the mode order, enabling customization of the FUC's topological properties. Here, two cases are considered: (i) both the acoustic and electromagnetic bandgaps are topologically trivial, and (ii) the acoustic bandgap is topologically trivial while the electromagnetic one is nontrivial.

In the first case, both the acoustic and electromagnetic bandgaps are set to topologically trivial phase. The band structures of the designed FUCs are presented in Fig.~\ref{fig16}. The center frequencies of the acoustic bandgaps approximately align with the specified reference frequencies, demonstrating successful bandgap positioning. However, the bandgaps of dual waves are relatively narrow, which poses challenges in aligning them with the PxC CUC. The same topological characteristics of both wave types restrict the bandgap width.

In the second case, the acoustic bandgap is topologically trivial and the electromagnetic one is nontrivial. For different acoustic reference frequencies, the FUCs and their band structures are shown in Fig.~\ref{fig17}. The bandgap widths are significantly larger when the topological properties of different wave types are opposite, compared to those in the first case. Besides, it is found that a higher reference frequency leads to the PxC UC with a lower filling ratio.	






\bibliographystyle{elsarticle-num-names}
\bibliography{references}

\end{document}